\documentclass[12pt,preprint]{aastex}

\begin{document}

\title{Long-term Optical Studies of the Be/X-ray Binary RX~J0440.9+4431/LS~V+44~17}

\author{Jingzhi Yan\altaffilmark{1}, Peng Zhang\altaffilmark{1,2}, Wei Liu\altaffilmark{1,2}, and Qingzhong Liu\altaffilmark{1}}

\altaffiltext{1}{Key Laboratory of Dark Matter and Space Astronomy, Purple Mountain Observatory, Chinese Academy of Sciences, Nanjing 210008, China}
\altaffiltext{2}{University of Chinese Academy of Sciences, Yuquan Road 19, Beijing 100049, China}
\email{jzyan@pmo.ac.cn}

\begin{abstract}
We present the spectroscopic and photometric observations on the Be/X-ray binary RX~J0440.9+4431 from 2001 to 2014. The short-term and long-term variability of the H$\alpha$ line profile indicates that one-armed global oscillations existed in the circumstellar disk. Several positive and negative correlations between the $V$-band brightness and the H$\alpha$ intensity were found from the long-term photometric and spectroscopic observations. We suggest that the monotonic increase of the $V$-band brightness and the H$\alpha$ brightness between our 2005-2007 observations might be the result of a continuous mass ejection from the central Be star, while the negative correlation in 2007-2010 should be caused by the cessation of mass loss from the Be star just before the decline in $V$-band brightness began (around our 2007 observations).  With the extension of the ejection material, the largest circumstellar disk during the last two decades has been observed in our 2010 observations with an equivalent width of $\sim$~$-12.88$ \AA, which corresponds to a circumstellar disk with a size of 12.9 times the radius of the central Be star. Three consecutive X-ray outbursts peaking around MJD~55293, 55444, and 55591 might be connected with the largest circumstellar disk around the Be star. We also use the orbital motion of the neutron star as a probe to constrain the structure of the circumstellar disk and estimate the eccentricity of the binary system to be $\ge$ 0.4. After three years of the H$\alpha$ intensity decline after the X-ray outbursts, a new circumstellar disk was being formed around the Be star after our 2013 observations.

\end{abstract}


\section{Introduction}

Be/X-ray binaries are a major subclass of the high mass X-ray binaries in our Galaxy \citep{Liu06}. The Be/X-ray binary system is composed of a Be star and a neutron star in an eccentric orbit \citep{Reig11}. The Be stars are non-supergiant rapid-rotating B-type star, showed the Balmer lines in emission at least once in its life \citep{Porter03}. Until the beginning of 2014, no Be-black hole system has been observed in the Be/X-ray binaries \citep{Belczynski2009,Zhang04}. The first Be-black hole system was first found in MWC~656 by \citet{Casares2014} in 2014 January. The optical emission in the Be/X-ray binary is mainly dominated by the Be star, while the X-ray emission reflects the physical condition in the vicinity of the neutron star. There are two different X-ray outbursts in the Be/X-ray binaries, namely Type I X-ray outburst and Type II X-ray outburst. Type I X-ray outbursts usually occurred around the periastron point of the neutron star, while the Type II X-ray outburst could happen at any orbital phase, which might be connected with the warping of the the outermost part of the circumstellar disk \citep{Okazaki2013}.

RX~J0440.9+4431 was first discovered as a Be/X-ray binary from the ROSAT galactic plane survey data \citep{Motch1997}. Its optical counterpart is the variable star BSD~24-491 = LS~V~+44~17 = VES~826, with a spectral type of B0.2Ve \citep{Reig2005}. An X-ray pulsation of 202.5$\pm$0.5s was identified from the RXTE/PCA light curves of RX~J0440.9+4431 in the energy band of 3-20 keV and its X-ray properties was similar to those of the other persistent Be X-ray binary pulsars, 4U~0352+309/X Per and RX J0146.9+6121/LS~I~+61~235 \citep{Reig1999}. An X-ray outburst from RX~J0440.9+4431 was observed between 2010 March 26 and 2010 April 15 by MAXI \citep{Morii2010} and the following two small X-ray flares were also detected by Swift/BAT \citep{Krivinos2010}. An orbital period of 150.0$\pm$0.2 days was estimated from three consecutive X-ray outbursts in the Swift/BAT light curve of RX~J0440.9+4431 by \citet{Ferrigno2013}.  A $\sim$32~keV cyclotron resonant scattering feature was discovered in the source spectrum and the magnetic field strength of the neutron
star was estimated as B $\simeq$ 3.2 $\times$ $10^{12}$ G \citep{Tsygankov2012}. The pulse profile has a sinusoidal-like single-peaked shape \citep{Tsygankov2012} and a narrow dip structure was clearly seen in the pulse profile of the low energy bands \citep{Usui2012}.

The spectroscopic and photometric observations indicate that RX~J0440.9+4431 showed a  moderately reddened, E(B-V) = 0.65$\pm$0.05, which corresponds to a distance of $\sim$ 3.3$\pm$0.5 kpc for a B0.2V star \citep{Reig2005}. The H$\alpha$ line had a double-peaked profile, varying from a symmetric structure to completely distorted on one side. The equivalent width (EW) of the H$\alpha$ line and the near infrared magnitudes showed a correlation and the variability of the H$\alpha$ line was attributed to the physical changes in the Be star's circumstellar disc \citep{Reig2005}.

In this work, we present our optical spectroscopic observations from 2001 to 2014 and the synchronous photometric observations from 2007 to 2014 of RX~J0440.9+4431.

\section[]{Observations}
\subsection{Optical Spectroscopy}

Most of our spectra of RX~J0440.9+4431 were obtained with the 2.16m telescope at Xinglong Station of National Astronomical Observatories. The optical spectroscopy with an intermediate resolution  was made with a CCD grating spectrograph at the Cassegrain focus of the telescope. We took the red spectra covering from 5500 to 6700\,{\AA}. Sometimes low-resolution spectra (covering from 4300 to 6700\,{\AA}) were also obtained. In 2012 March and 2013 November, we also carried out the spectroscopic observations with the Yunnan Faint Object Spectrograph and Camera (YFOSC) instrument of the Lijiang 2.4m telescope in Yunnan Astronomical Observatory. The Grism \#8 was used with a resolution of 1.47\,{\AA} pixel$^{-1}$, with a spectral range from 5050 to 9750\,{\AA}. The journal of our observations is summarised in
Table~\ref{table:log}, including observational date, exposure time, Modified Julian Date (MJD) and spectral resolution, respectively. All spectroscopic data were reduced with the IRAF\footnote{IRAF is distributed by NOAO, which is
operated by the Association of Universities for Research in Astronomy, Inc., under cooperation with
the National Science Foundation.} package. They were bias-subtracted, flat-field corrected, and
had cosmic rays removed. Helium-argon spectra were taken in order to obtain the pixel-wavelength
relations.

The EWs of the H$\alpha$ line have been measured by selecting a continuum point on each side of the line and integrating the flux relative to the straight line between the two points using the procedures available in IRAF. The measurements were repeated five times for each spectrum and the error is estimated from the distribution of these values. The typical error for the measurements is within 1\%, arising due to the subjective selection of the continuum. The results of H$\alpha$ EWs are listed in Table~\ref{table:log} and plotted in Figure~\ref{figure:all}. The published data of H$\alpha$ \citep[adopted from][]{Reig2005}, are also plotted in Figure~\ref{figure:all} with open symbols.

\subsection{Optical Photometry}

Since 2007, we performed systematic photometric observations on a sample of X-ray binaries with the 100~cm Education and Science Telescope (EST) and the 80~cm Tsinghua-NAOC Telescope (TNT) at Xinglong Station of NAOC. The EST, manufactured by EOS Technologies, is an altazimuth-mounted reflector with Nasmyth foci at a focal ratio of $f/8$. TNT is an equatorial-mounted Cassegrain system with a focal of $f/10$, made by AstroOptik, funded by Tsinghua University in 2002 and jointly operated with NAOC \citep{Huang2012}. Both telescopes are equipped with the same type of Princeton Instrument 1340$\times$1300 thin back-illuminated CCD. The CCD cameras use standard Johnson-Cousins $UBVRI$ filters made by Custom Scientific.

\begin{figure}

\centerline{\includegraphics[width=12cm]{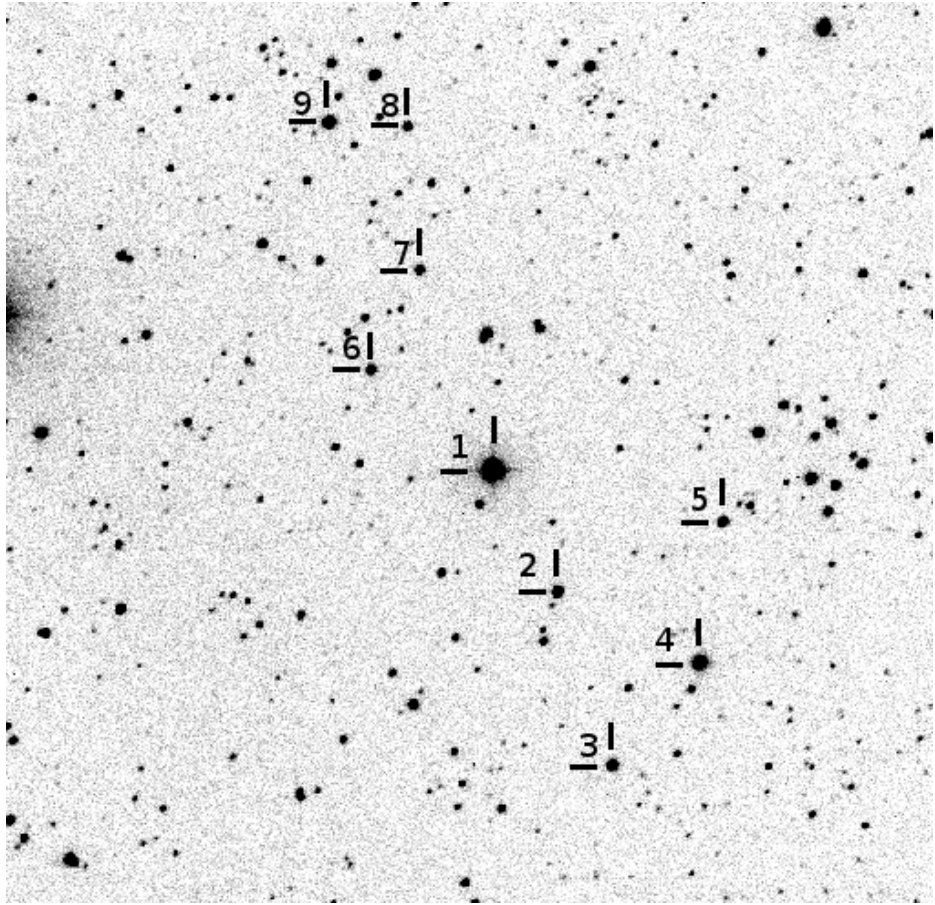}}

    \caption{The field stars of RX~J0440.9+4431 in $V$-band.  RX~J0440.9+4431 is marked with \#1 and eight reference stars are marked with \#2 to \#8. North is at the bottom and East to the left.}
    \label{figure:field}
\end{figure}

\begin{figure}

\centerline{\includegraphics[width=13cm]{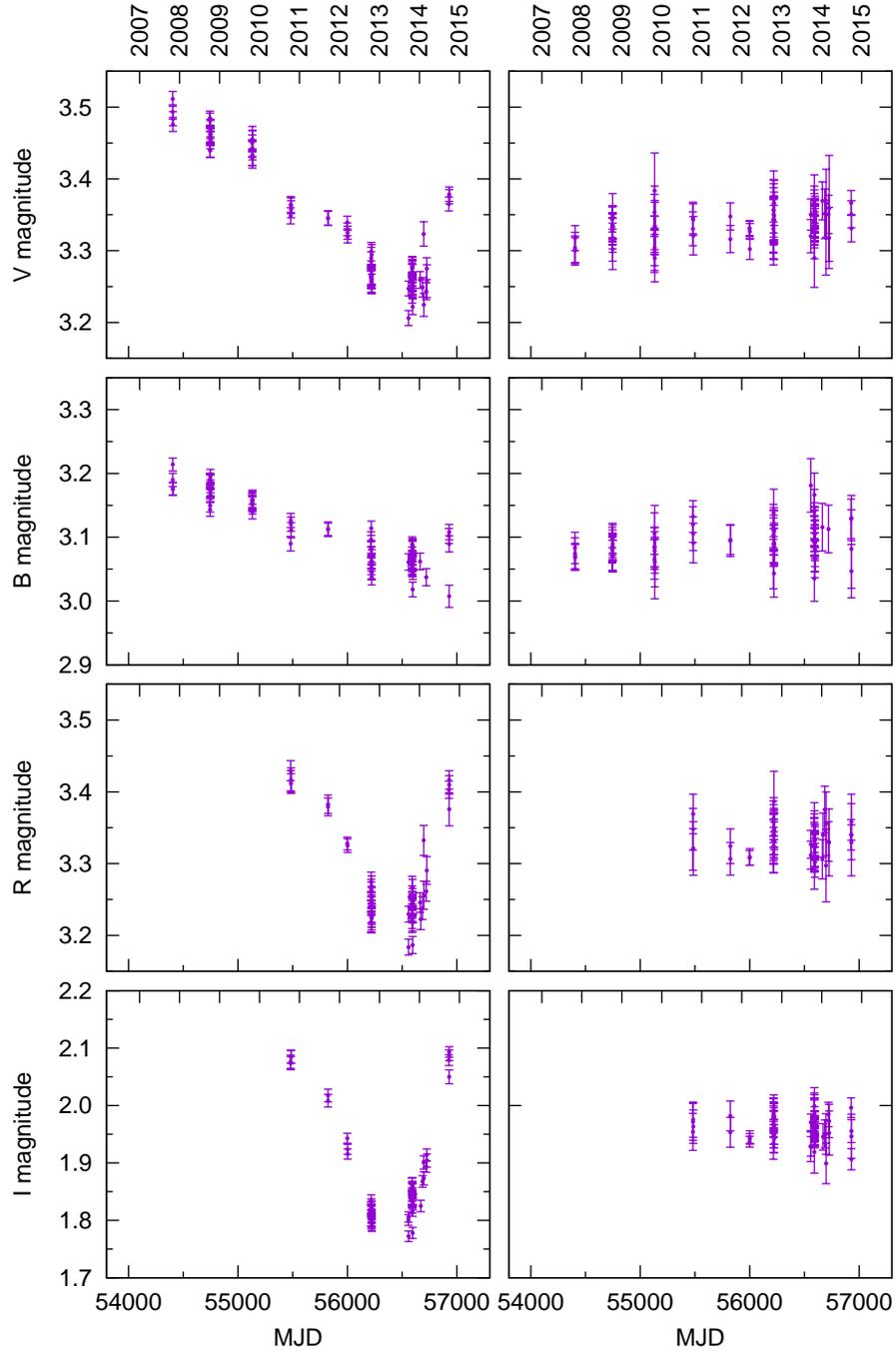}}

    \caption{The $BVRI$ differential photometry of RX~J0440.9+4431. The offsets of 4.2 mag, 4.4 mag, 4.1 mag and 3.8 mag, are added to the $V$, $B$, $R$, and $I$ differential light curves of the reference star \#8 and plotted them in the right panels. The first day of each year is also marked at the top of each panel with a format of YYYY.}
    \label{figure:diff}
\end{figure}

The photometric data reduction was performed using standard routines and aperture photometry packages in IRAF, including bias subtraction and flat-field correction. In order to study the variation of the optical brightness, we selected eight reference stars in the field of view (see Fig.~\ref{figure:field}) to derive the differential magnitude of RX~J0440.9+4431. We follow the algorithm of \citet{Broeg2005} to compute an artificial companion star using all the eight reference stars with different weights. The $BVRI$ differential magnitudes and their errors are listed in Table~\ref{table:logphot} and plotted in Figure~\ref{figure:diff} together with the light curves of the reference star \#8 in the right panel. In order to plot the same level with the target star, we add some offsets to the $BVRI$ magnitudes of the reference star \#8 in Figure~\ref{figure:diff}.

\begin{figure}

\centerline{\includegraphics[width=8cm]{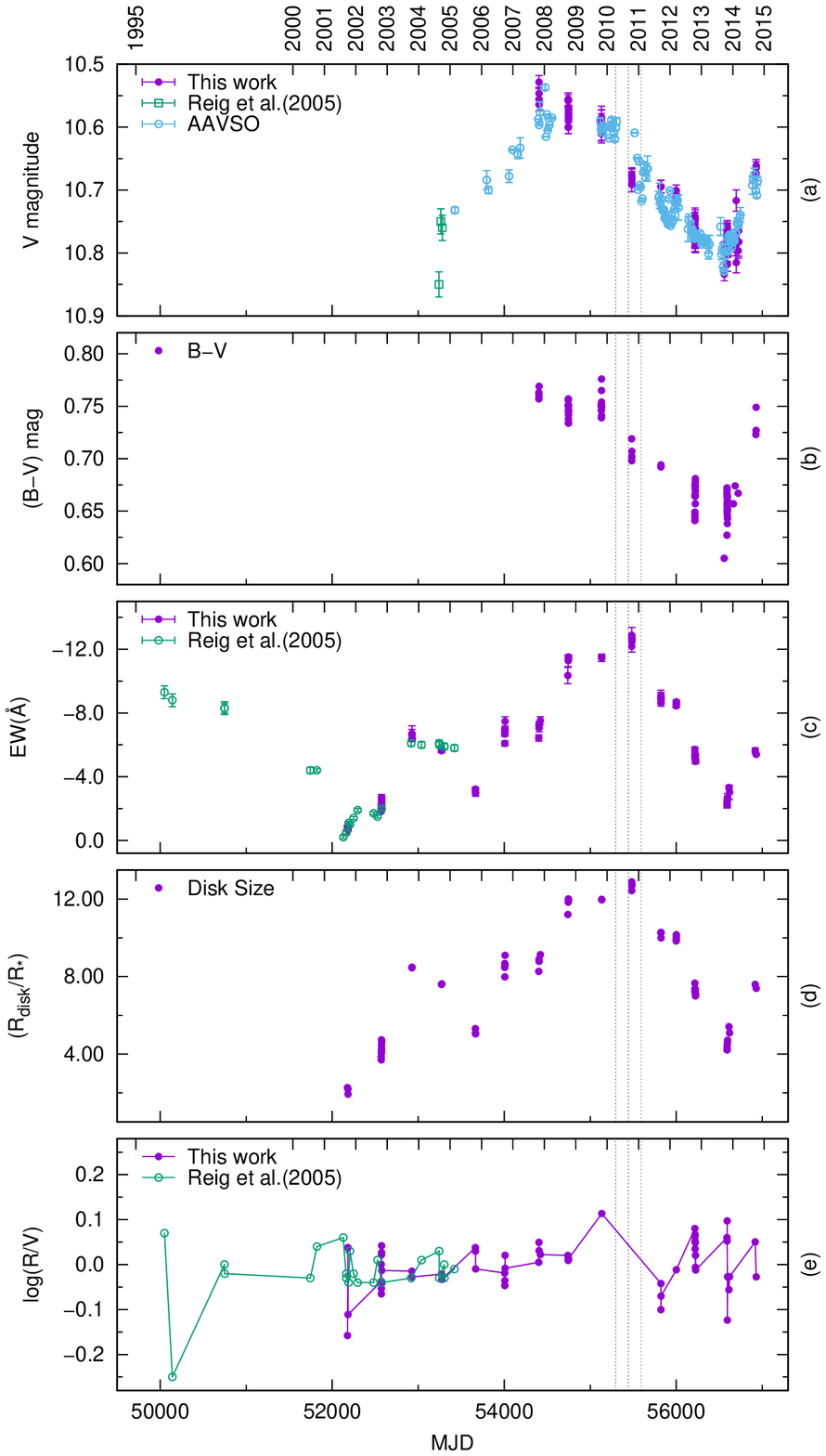}}

    \caption{(a). $V$-band brightness between our 2007 and 2014 observations marked with filled circles. The public data adopted from \citet{Reig2005} and $AAVSO$ are marked with open symbols; (b). ($B-V$) colour index between our 2007 and 2014 observations; (c). the EWs of the H$\alpha$ lines between our 2001 and 2014 observations marked with filled circles. The data adopted from \citet{Reig2005}  are marked with open symbols; (d). the H$\alpha$ emission region in the unit of the central star radius; (e). the logarithmic function of $V/R$ between our 2001 and 2014 observations marked with filled circles and the public data adopted form \citet{Reig2005} with open symbols. Three dashed lines are marked as the time of three consecutive X-ray outbursts, peaked around MJD~55293, 55444, and 55591, respectively. The first day of each year is also marked at the top of each panel with a format of YYYY.
    }
    \label{figure:all}
\end{figure}

In order to study the long-term optical brightness variability of the source, we also make use of the public optical photometric data from the international database of the American Association of Variable Star Observers (AAVSO\footnote{http://www.aavso.org/}). Only the photometric observations made with the Johnson $V$-band filter are used  and plotted in Figure~\ref{figure:all} with 10-day bins. Using the secondary standard star C5 (our reference star \#5 in Figure~\ref{figure:field}, with $\bar{V}$=14.625 mag and $\bar{B}$=15.665 mag) listed in \citet{Reig2015}, we also plot our $V$-band results with an offset of $\sim$ 14.05 mag in Figure~\ref{figure:all}. In addition, we also adopt the photometric data in \citet{Reig2005} and plot them in Figure~\ref{figure:all} with open symbols. The ($B-V$) colour of the source between our 2007 and 2014 observations is also plotted in Figure~\ref{figure:all}.

\subsection{X-ray Observations}
The Burst Alert Telescope (BAT) on board \emph{Swift} \citep{Barthelmy00} monitored RX~J0440.9+4431 in the hard X-ray energy band (15--50~keV) since 2005 February. Three consecutive X-ray outbursts were detected between 2010 March and 2011 February \citep{Ferrigno2013}. The positions of the peak time for three X-ray outbursts are marked with three dashed lines in Figure~\ref{figure:all}.

\section{Results}
\subsection{Long-term and short-term variability of the H$\alpha$ line}
\begin{figure}

\centerline{\includegraphics[width=10cm]{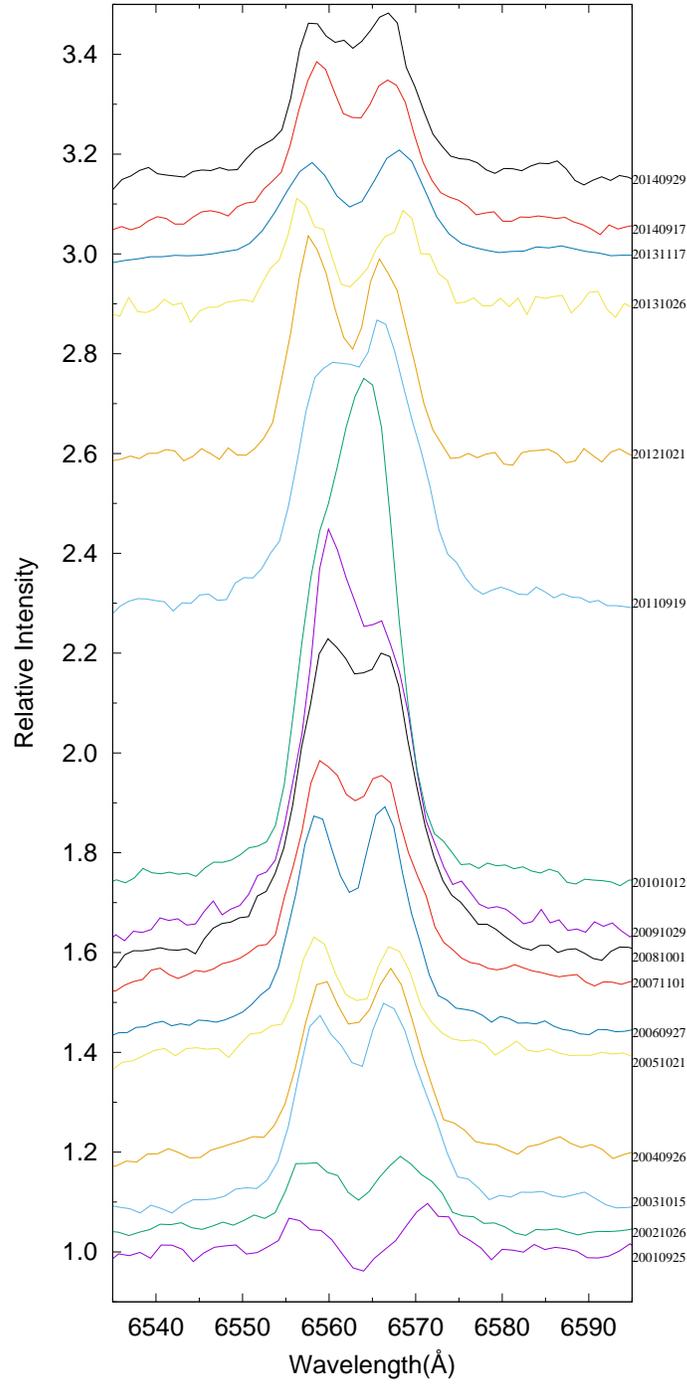}}

    \caption{The H$\alpha$ profiles selected from our 2001-2014 observations (separated in rectified intensity for clarity). Observational date is marked on the right part of each spectrum with a format of YYYYMMDD.}
     \label{figure:profile}
\end{figure}

The profile of H$\alpha$ lines showed subtle change during each observational run. We select one typical H$\alpha$ line from each run during our 2001--2014 observations and plot them in Figure~\ref{figure:profile}, which indicates that the H$\alpha$ profile of RX~J0440.9+4431 usually showed an obvious double-peaked profile, except the observations in 2010, when the H$\alpha$ line displayed a strong single-peaked structure. Symmetric double-peaked H$\alpha$ lines were observed during 2001 and 2002 observations. The central depression of the double-peaked H$\alpha$ line during our 2001 observations was below the stellar continuum, which was also reported in \citet{Reig2005}. The intensity of H$\alpha$ during the 2001 observations was in the lowest level during our 15-year observations. Its intensity showed  an obvious increase in the following two years. The H$\alpha$ intensity showed an abrupt decrease in 2005 and it had the same level as that in 2002. After that, the intensity of the H$\alpha$ line kept increasing in the following five years and reached a peak value with an averaged EW of $\sim$ $-12.9$~\AA~during our 2010 observations. It is interesting that asymmetric profiles with a strong red peak were observed during 2008 and 2009 observations, while only a nearly symmetric single-peaked H$\alpha$ line was observed during 2010.  After 2010, the intensity of the H$\alpha$ line showed a rapid decline and had an EW of $\sim$$-5$~\AA~during our 2012 October observations. The typical H$\alpha$ profile during 2011 was an asymmetric structure with a strong violet peak, while the profile during 2012 October with a stronger red peak. Figure~\ref{figure:all} indicates that the intensity of the H$\alpha$ line kept decreasing before our 2013 October observations and it had a faint nearly symmetric double-peaked profile in our 2013 October 26 spectrum. The H$\alpha$ intensity during our 2013 October observation was at a low emission level, which is similar to the level during our 2012 observations. One moth later, we observed a stronger H$\alpha$ line during our 2013 November observations. The H$\alpha$ intensity kept increasing during the following year and its emission during our 2014 observations returned to the level during 2012 observations. The asymmetric H$\alpha$ lines during our 2014 observations showed rapid changes, with a stronger violet peak in the 2014 September 17 spectrum, while a stronger red peak on 2014 September 29. 

\begin{figure}

\centerline{\includegraphics[width=10cm]{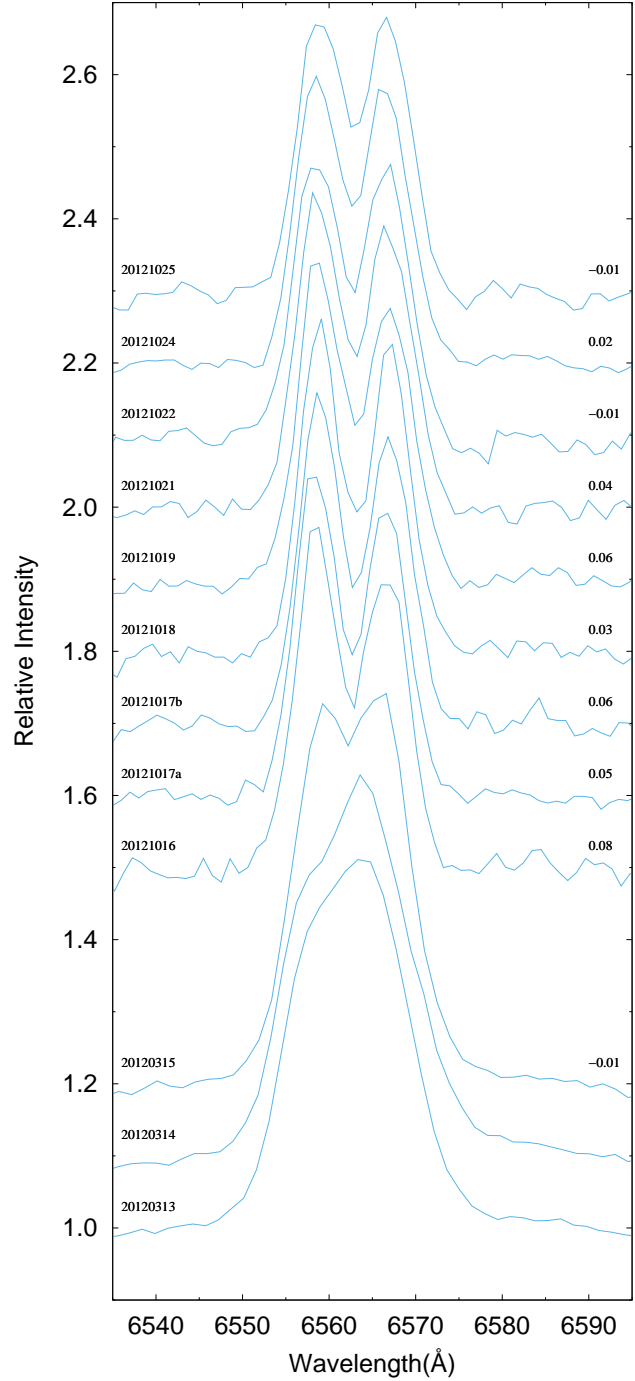}}

    \caption{The evolution of the H$\alpha$ profiles during our 2012 observations. The observational date and the $V/R$ ratio are marked on the left and right wings of each spectrum, respectively. }
      \label{figure:2012}

\end{figure}

Following the method of \citet{Reig2005}, we also define the $V/R$ ratio of the H$\alpha$ line as $V/R = (I(V)-I_c)/(I(R)-I_c)$, where $I(V)$, $I(R)$, and $I_c$ are the intensities of the violet peak, red peak and continuum, respectively. The $V/R$ ratios of the H$\alpha$ lines using logarithmic scales are listed in Table~\ref{table:log} and plotted in Figure~\ref{figure:all}, together with the data adopted from \citet{Reig2005} with open symbols. The $V/R$ ratio could change rapidly within each observational run. In the long run we can not see any periodic variation from the $V/R$ ratio curves.  
.
In order to study the short-term variability of the H$\alpha$ line, we plot the H$\alpha$  lines observed in 2012 March and October at Lijiang Observatory and Xinglong Station, respectively, in Figure~\ref{figure:2012}.  On the right region of each spectrum, it is marked with the value of $log(V/R)$. Due to the lower resolution of the spectra observed at Lijiang station, only an asymmetric profile with strong red peak was observed on 2012 March 13 and 14. A clearly double-peaked profile with a stronger red peak was observed on the spectrum of 2012 March 16. During the 2012 October observations, most of our spectra showed as an asymmetric profile with strong violet peak, except for the spectra observed on 2012 October 22 and 25  with a stronger red peak.

\subsection{The circumstellar disk size around the Be star}
The intensity of the H$\alpha$ emission line can be used as an indicator of disk size around the Be star. \citet{Hanuschik1989} found a correlation between the ratio of the H$\alpha$ double peak separation ($\Delta{v_{peak}}$) and the Be star projected rotational velocity $v{\rm sin}i$, and the H$\alpha$ EWs with a factor of 2 scatter given by

\begin{equation}
\label{euqation1}
log[\Delta{v_{peak}}/2v{\rm sin}i]=-0.32log(EW_{H\alpha})-0.20,
\end{equation}
where, $EW_{H\alpha}$ is in a unit of $-\AA$. Assuming a Keplerian circumstellar disk around the Be star, we can estimate the disk size using Huang's formula \citep{Huang1972} ,

\begin{equation}
\label{euqation2}
R_d/R_*=(2v{\rm sin}i/\Delta v_{peak})^2,
\end{equation}
where $R_d$ is the radius of the H$\alpha$ emission region, $R_*$ the radius of the central Be star. Following Equation(3) in \citet{Coe2015}, we can estimate the disk size around the Be star using the H$\alpha$ EWs with the following equation

\begin{equation}
\label{euqation3}
log\sqrt{R_*/R_d}=-0.32log(EW_{H\alpha})-0.20.
\end{equation}

The H$\alpha$ emission regions are calculated by Equation~\ref{euqation3} in the unit of the central star radius $R_*$ and the results are listed in Table~\ref{table:log} and showed in Figure~\ref{figure:all}. The maximum circumstellar disk with a radius of $\sim$ 12.9$R_*$ corresponds to the strongest H$\alpha$ emission during our 2010 observations.

\subsection{The long-term evolution of  the $BVRI$-band brightness}
Our photometric observational results of RX~J0440.9+4431 with $BVRI$ filters are plotted in Figure~\ref{figure:diff}. Only $BV$-band data are available between the 2007 and 2009 observations. After 2010, we monitored the source with the $BVRI$ filters. Figure~\ref{figure:diff} indicates that the $V$-band brightness of RX~J0440.9+4431 showed a continuous decrease between our 2007 and 
2013 observations and it had a minimum brightness around MJD~56595. The source faded by about 0.3 magnitude during the six years. After our 2013 October observations, a rapid increase was observed in the $V$-band photometry and the system brightened by about 0.15 magnitude within nearly one year.  The $B$-band observations showed the same evolutionary way as the $V$-band observations, but the $B$-band magnitudes decreased more slowly between our 2007 and 2013 observations with an amplitude of $\sim$~0.2 magnitude. The brightness of the source in $RI$-band observations has a more rapid variation. After the system reached a minimum brightness during our 2013 observations, the $R$ and $I$ brightness increased about 0.2 and 0.3 magnitude, respectively, within about one year. 

The ($B-V$) colour index kept decline on average between our 2007 and 2013 observations, which indicates that the spectrum of the Be star became bluer and bluer. The bluer spectrum means that the contribution of the disk emission decreased by the depletion of the inner part of the disk once the mass supply from Be star stopped around our 2007 observations (we will discuss this in the next section). With the diffusion of the circumstellar material around the Be star out into the outer space, its spectrum became bluer and bluer. When the brightness began to increase after the 2013 October observations, a reddening of ($B-V$) was also observed simultaneously. This means that a new circumstellar disk was being formed around the Be star. 

\subsection{The correlation between the H$\alpha$ intensity and the $V$-band brightness}

Due to the lack of V-band data before MJD $\sim$ 53250, we can only study the correlation between the H$\alpha$ emission and the $V$-band brightness after that time. Figure~\ref{figure:all} indicates that the $V$-band brightness increased continuously between MJD~$\sim$~53250 and $\sim$~54400 with an amplitude of 0.2 magnitudes on average. During this period, the intensity of the H$\alpha$ line showed a complicated variability. Subtle changes were observed between our 2003 and 2004 H$\alpha$ intensity, changing from $\sim$$-6.5$\AA~to $\sim$$-5.8$\AA. The H$\alpha$ line had a low emission level with an averaged EW of $-2.9$\AA ~during our 2005 October observations. The optical brightness had an obvious increase after our 2005 observations. The intensity of the H$\alpha$ line also became stronger during our 2006 September observations with an averaged EW of $\sim$$-7.0$\AA.  The H$\alpha$ emission during our 2007 November observations had the same level as that during 2006, while the synchronous photometric observations indicates that the $V$-band brightness of the system had the highest level in the last two decades.  After that, the $V$-band brightness showed a continuous decline between our 2007 and 2013 September observations, with an amplitude of $\sim$~0.3 mag, which has been discussed above. Unlike the evolutionary behaviour of the $V$-band brightness, the intensity of the H$\alpha$ line showed an abrupt increase during our 2008 September observations. Its intensity kept increasing in the following two years and reached a maximum with an EW of $\sim$ $-12.9$\AA~during our 2010 October observations, which was the strongest H$\alpha$ emission that has been reported in the literature. After 2010, the H$\alpha$ intensity showed a continuously rapid decline and had a minimum EW of $\sim$ $-2.25$\AA  ~during our 2013 October observations, changing with an amplitude of $\sim$~10.35\AA~within three years of observations. After they reached their minima around our 2013 October observations, both of the $V$-band brightness and the intensity of the H$\alpha$ line showed a rapid increase. The $V$-band magnitude brightened about $\sim$~0.17 mag and the intensity of the H$\alpha$ line was in a level of $\sim$$-5.6$\AA~during our 2014 September observations.

\section{Discussions}
We present the long-term optical spectroscopic observations (2001--2014) and the synchronous photometric observations (2007-2014) on the Be/X-ray binary RX~J0440.9+4431. Combining with the public data, we studied the short-term and long-term activities of the system in optical.

The profile of the H$\alpha$ line displays short-term and long-term V/R variability, which has been discussed by \citet{Reig2005}. The cyclic V/R changes are observed in many single Be stars \citep{Porter03} and they might be connected with the density distributions of the material in the circumstellar disk. The popular theoretical explanations for the V/R variations is the one-armed oscillation model \citep{Okazaki91}. The V/R variability in Be/X-ray binaries might be also connected with the tidal interaction between the neutron star and the circumstellar disk around the Be star \citep{Oktariani2009}.

As we discussed in the last section, the EW of the H$\alpha$ line can be used to estimate the disk radii around the Be star. The disk loss event occurred in early 2001 had been discussed in \citet{Reig2005}.  A new disk was reforming during our 2001 and 2002 observations, and it reached a relative stable state during the following two years of observations. With the dissipation of the material, a relatively small disk was observed during our 2005 observations. After then both the optical brightness and the intensity of the H$\alpha$ line showed rapid increase, implying that a larger disk had been formed during our 2006 observations. It is interesting that the emission of the H$\alpha$ line during our 2007 observations had the same level as that during our 2006 observations, while the $V$-band brightness during 2007 was in the highest state for all the observations with $\sim$10.5 magnitudes.  Even more interesting is that the intensity of the H$\alpha$ line still kept increasing when the optical brightness declined after our 2007 observations. The circumstellar disk had the largest radii during our 2010 observations. For the next three years, the disk was in another rapid dissipation phase. The circumstellar disk was in another low state during our 2013 October observations, which is similar to our 2005 observations.  A new circumstellar disk was being reformed after 2013 October. 

The monotonic increase of the system brightness and the intensity of the H$\alpha$ line observed between our 2005 and 2007 observations indicates that a larger and denser circumstellar disk was being formed during this period, which might be the result of a continuous mass ejection from the central Be star. The negative correlation between the $V$-band brightness and the H$\alpha$ emission during our 2007-2010 observations (see Figure~\ref{figure:all}) should have a different physical mechanism with the positive one. The similar negative correlation between the optical brightness and the intensity of the H$\alpha$ line has been observed in other Be/X-ray binaries, such as A~0535+26 \citep{Yan2012a}, 4U~0115+63 \citep{Reig07}, MXB~0656-072 \citep{Yan2012b}, and SAX~J2103.5+4545 \citep{Camero2014} . As discussed in our previous papers \citep{Yan2012a,Yan2012b}, only the inner parter of the circumstellar disk has a major contribution to the variability of the $V$-band brightness, while the H$\alpha$ emission comes from the nearly entire disk \citep{Slettebak1992,Stee1998,Carciofi2011}. The negative correlation between the $V$-band brightness and the H$\alpha$ emission during our 2007-2010 observations might be explained as the result of the cessation mass supply from the Be star in the system. \citet{Rivinius2001} and \citet{Meilland2006} suggested that a low-density region would be developed after several months of the mass ejection from the surface of the Be star, basing on the observational results and the theoretical calculations, respectively. The mass ejection stopped just before the decline in the V-band brightness began (around our 2007 observations). This cessation of mass supply from the central Be star caused the depletion in the innermost part of the circumstellar disk and the fading in the V-band brightness. In another hand, the outer part of the disk was still growing. In other words, the viscous diffusion was still going on in the outer part. This could explain the increase of the H$\alpha$ intensity when the $V$-band brightness showed a decrease after our 2007 observations. 

 When the outer part of the disk moved near the periastron passage of the neutron star, accretion rate onto the neutron star would increase and an Type I X-ray outburst could be triggered.  Three consecutive X-ray outbursts, peaking around MJD~55293, 55444, and 55591, respectively, were observed in Be/X-ray binary RX~J0440.9+4431. On the other hand, the circumstellar disk would be truncated by the orbital motion of the neutron star \citep{Okazaki01}. Figure~\ref{figure:all} indicates that the intensity of the H$\alpha$ line showed a rapid decline after our 2010 observations. Its EW changes from $-12.9$~\AA~to $-2.3$~\AA~within three years. After three years of the H$\alpha$ intensity decline, the Be star in the system was in another active phase and a new circumstellar disk was being formed.

In Be/X-ray binaries, the neutron star can be used as a probe to constrain the physical structure of the circumstellar disk around the Be star. The three consecutive X-ray outbursts of RX~J0440.9+4431 took place during the time when the system showed a strong H$\alpha$ emission. During the X-ray outburst, the size of the circumstellar disk should be larger than the Roche lobe radius of the Be star at the periastron. The Roche lobe radius of a Be star $R_L$ is given approximately by
\begin{equation}
R_L=D \frac{0.49 q^{2/3}}{0.69 q^{2/3}+ln(1+q^{1/3})}
\end{equation}
\citep{Eggleton1983}. Here $q$ is the mass ratio $M_*/M_X$, where $M_*$ and $M_X$ are the mass of the Be star and the neutron star, respectively, and $D$ is the instantaneous distance between the binary. We can also estimate the semi-major axis of the binary system with Kepler's third law,  $a^3=P_{\mathrm{orb}}\times G(M_*+M_{\mathrm{X}})/4\pi^2$. For a B0.2Ve star \citep{Reig2005}, we take its mass and radius as $M_*$ = 18 $M_{\odot}$ and $R_*$ = 8 $R_{\odot}$, respectively \citep{Negueruela2001}. The canonical mass of a neutron star is fixed to $M_{\mathrm{X}}=1.4\ M_{\sun}$. With an orbital period of $P_{\mathrm{orb}}$= 150d \citep{Ferrigno2013}, the semi-major axis is derived as $a\sim 319.4\ R_{\sun}\sim 40\ R_*$ and the Roche lobe size $R_L$=0.54$D$. When the system showed the X-ray outburst, the largest circumstellar disk with a size of $R_d$$\sim$ 12.9 $R_*$ during our 2010 observations should larger than the Roche lobe radius of the Be star at the periastron. With these parameters, we can estimate the eccentricity of the binary orbit by $R_L$(periastron)=0.54$a$(1-$e$) $\le$ $R_d(max)$. Therefore, the Be/X-ray binary RX~J0440.9+4431 should have an eccentric orbit with $e$ $\ge$ 0.4. Actually, the Be disk is not always circular. Sometimes, a Be disk becomes elongated. If the Be disk was elongated toward the periastron, the maximum disk size to be used in the above condition would be larger than the observed maximum H$\alpha$ emission radius. 

\section{Conclusions}

In this work, we report our optical spectroscopic and photometric observations on the Be/X-ray binary RX~J0440.9+4431 between 2001 and 2014. The profile of the H$\alpha$ line showed long-term and short-term V/R variabilities, which might be connected with the one-armed oscillation in the disk. The positive and negative correlations between the optical brightness and the intensity of the H$\alpha$ line were observed in the system. The monotonic increase of the $V$-band brightness and the H$\alpha$ brightness between our 2005-2007 observations might be the result of a continuous mass ejection from the central Be star. The negative correlation between the $V$-band optical brightness and the intensity of the H$\alpha$ line between our 2007 and 2010 observations could be explained as the result of the cessation of mass supply from the central Be star around our 2007 observations. With the stop of the mass supply from the Be star, a low density region would be formed in the inner part of the disk and the fading in the $V$-band brightness was resulted in. But, the viscous diffusion was still going on in the outer part and the intensity of the H$\alpha$ line kept increasing when the brightness showed a decline. With the outward motion of the ejected material, the largest circumstellar disk and the strongest H$\alpha$ emission line during the last 20 years, with an EW of $\sim$ $-12.9$\AA, ~was observed during our 2010 observations, which trigged three consecutive X-ray outbursts between 2010 March and 2011 February. After three years of the H$\alpha$ decline, a new circumstellar disk was being formed after our 2013 observations. As a probe of the circumstellar disk structure, we use the orbital motion of the neutron star to estimate the eccentricity of the system as $e$ $\ge$ 0.4.

\acknowledgements We thank the anonymous referee for his/her valuable comments to our manuscript. This work was supported in part by the 973 Program of China under grant 2013CB837000; the National Natural Science of China under grants 11273064, 11433009 and 11573071; and the Strategic Priority Research Program of CAS (under grant number XDB09000000). This work was partially Supported by the Open Project Program of the Key Laboratory of Optical Astronomy, National Astronomical Observatories, Chinese Academy of Sciences.  We also acknowledge the support of the staff of the Lijiang 2.4m telescope. Funding for the telescope has been provided by CAS and the People's Government of Yunnan Province.

\begin{deluxetable}{ccccccc}
\tablecolumns{7} \tablecaption{Summary of the spectroscopic observations of LSV+44~17 between our 2001 and 2014 observations.}
\tablewidth{0pc}
\tablehead{\colhead{Date} & \colhead{Exposure} & \colhead{MJD} & \colhead{Dispersion} &\colhead{EW(H$\alpha$) } & \colhead{log(V/R)$^{a}$}  & \colhead{$R_d/R_*$} \\
\colhead{(YYYY-MM-DD) }&\colhead{(s)} & \colhead{} & \colhead{(\AA/pixels)} &\colhead{(--\AA)} &\colhead{} & \colhead{}}  \startdata
2001-09-25 & 1000 & 52177.828 & 1.20 & 0.85 $\pm$ 0.09 & -0.16 & 2.3  \\
2001-09-30 & 1000 & 52182.822 & 1.20 & 0.81 $\pm$ 0.05  & 0.04  & 2.2 \\
2001-10-01 & 1000 & 52183.825 & 1.20 & 0.66 $\pm$ 0.07  & -0.11 & 1.9\\
2002-10-23 & 1000 & 52570.776 & 1.20 & 1.83 $\pm$ 0.03   & -0.04 & 3.7\\
2002-10-23 & 1200 & 52570.790 & 1.20 & 1.95 $\pm$ 0.12   & 0.00  & 3.8\\
2002-10-24 & 1000 & 52571.856 & 1.20 & 2.11 $\pm$ 0.15   & -0.05  & 4.0\\
2002-10-24 & 1000 & 52571.879 & 1.20 & 2.18 $\pm$ 0.10   & -0.06  & 4.1\\
2002-10-26 & 1000 & 52573.750 & 1.20 & 2.46 $\pm$ 0.15   & -0.04   & 4.5\\
2002-10-26 & 1200 & 52573.763 & 1.20 & 2.69 $\pm$ 0.19   & 0.02   &4.7\\
2002-10-26 & 1200 & 52573.845 & 1.20 & 2.16 $\pm$ 0.05   & 0.03   & 4.1\\
2002-10-27 & 1200 & 52574.821 & 1.20 & 2.31 $\pm$ 0.04   & 0.04   &4.3\\
2002-10-28 & 1200 & 52575.785 & 1.20 & 2.62 $\pm$ 0.15   & -0.01  & 4.6\\
2003-10-14 & 1200 & 52926.796 & 1.20 & 6.66 $\pm$ 0.30   &-0.01   & 8.4\\
2003-10-15 & 800 & 52927.842 & 1.20 & 6.70 $\pm$ 0.49     & -0.03   &8.5\\
2003-10-16 & 100 & 52928.730 & 2.45 & 6.42 $\pm$ 0.17     & --         & --\\
2003-10-16 & 200 & 52928.732 & 2.45 & 6.38 $\pm$ 0.06     & --         & --\\
2004-09-22 & 900 & 53270.869 & 1.20 & 5.63 $\pm$ 0.12     &-0.02    & 7.6\\
2004-09-26 & 600 & 53274.811 & 1.20 & 5.67 $\pm$ 0.08     &-0.03     & 7.6\\
2005-10-21 & 900 & 53664.862 & 1.20 & 3.22 $\pm$ 0.13     & 0.04     & 5.3\\
2005-10-23 & 300 & 53666.740 & 2.45 & 3.01 $\pm$ 0.21     & 0.03     & 5.1\\
2005-10-24 & 900 & 53667.804 & 1.20 & 2.97 $\pm$ 0.19     & -0.01    & 5.0\\
2006-09-27 & 1200 & 54005.829 & 1.02 & 6.68 $\pm$ 0.15   & -0.02    & 8.5\\
2006-09-28 & 900 & 54006.786 & 1.02 & 6.09 $\pm$ 0.16     & -0.05    & 8.0\\
2006-09-29 & 900 & 54007.786 & 1.02 & 6.94 $\pm$ 0.14     & -0.04    & 8.7\\
2006-10-01 & 1200 & 54009.822 & 1.02 & 6.84 $\pm$ 0.19   & 0.02     & 8.6\\
2006-10-02 & 900 & 54010.816 & 1.02 & 7.47 $\pm$ 0.29     & -0.01    & 9.1\\
2007-10-28 & 900 & 54401.879 & 1.02 & 6.43 $\pm$ 0.20     &0.00      & 8.3\\
2007-10-31 & 900 & 54404.900 & 1.02 & 7.22 $\pm$ 0.21     &0.05      & 8.9\\
2007-11-01 & 600 & 54405.825 & 1.02 & 7.07 $\pm$ 0.26     & 0.03     & 8.8\\
2007-11-16 & 600 & 54420.721 & 1.02 & 7.52 $\pm$ 0.24     & 0.02     & 9.1\\
2008-10-01 & 1200 & 54740.761 & 1.02 & 10.35 $\pm$ 0.51 &0.02      & 11.2\\
2008-10-05 & 1200 & 54744.783 & 1.02 & 11.28 $\pm$ 0.36  & 0.01    & 11.8\\
2008-10-06 & 1200 & 54745.783 & 1.02 & 11.52 $\pm$ 0.13   & 0.01   & 12.0\\
2008-10-09 & 1200 & 54748.814 & 1.02 & 11.48 $\pm$ 0.13   &0.01    & 12.0\\
2009-10-28 & 900 & 55132.841 & 1.02 & 11.50 $\pm$ 0.12     & single &12.0\\
2009-10-29 & 900 & 55133.808 & 1.02 & 11.46 $\pm$ 0.23     &0.11     & 12.0\\
2010-10-12 & 1200 & 55481.855 & 1.02 & 12.18 $\pm$ 0.36   &single   & 12.4\\
2010-10-13 & 600 & 55482.861 & 1.02 & 12.88 $\pm$ 0.49     & single  & 12.9\\
2010-10-14 & 1200 & 55483.767 & 1.02 & 12.72 $\pm$ 0.12    & single  & 12.8\\
2010-10-15 & 1200 & 55484.809 & 1.02 & 12.59 $\pm$ 0.12    &single    & 12.7\\
2011-09-17 & 300 & 55821.814 & 1.02 & 8.88 $\pm$ 0.31        &single    & 10.2\\
2011-09-18 & 600 & 55822.790 & 1.02 & 9.01 $\pm$ 0.41        &-0.04     & 10.2\\
2011-09-18 & 600 & 55822.830 & 1.02 & 9.05 $\pm$ 0.24        &-0.10     & 10.3\\
2011-09-19 & 600 & 55823.805 & 1.02 & 8.65 $\pm$ 0.22        &-0.07     & 10.0\\
2012-03-13$^{b}$ & 600 & 55999.600 & 1.47 & 8.44 $\pm$ 0.12 & --       & 9.8\\
2012-03-14$^{b}$ & 500 & 56000.582 & 1.47 & 8.70 $\pm$ 0.11 & --        & 10.0\\
2012-03-15$^{b}$ & 300 & 56001.595 & 1.47 & 8.54 $\pm$ 0.16 & -0.01  & 9.9\\
2012-10-16 & 600 & 56216.847 & 1.02 & 5.71 $\pm$ 0.16           & 0.08   & 7.7\\
2012-10-17 & 600 & 56217.773 & 1.02 & 5.35 $\pm$ 0.08           & 0.05   & 7.3\\
2012-10-17 & 600 & 56217.887 & 1.02 & 5.23 $\pm$ 0.11            & 0.06   & 7.2\\
2012-10-18 & 600 & 56218.779 & 1.02 & 5.17 $\pm$ 0.10            & 0.03   & 7.2\\
2012-10-19 & 600 & 56219.776 & 1.02 & 5.38 $\pm$ 0.04            &0.06    & 7.4\\
2012-10-21 & 1200 & 56221.863 & 1.02 & 5.33 $\pm$ 0.11          & 0.04    &7.3\\
2012-10-22 & 600 & 56222.858 & 1.02 & 4.96 $\pm$ 0.15            &-0.01   & 7.0\\
2012-10-24 & 1200 & 56224.803 & 1.02 & 5.06 $\pm$ 0.12          &0.02    & 7.1\\
2012-10-25 & 1200 & 56225.804 & 1.02 & 4.98 $\pm$ 0.09          &-0.01   & 7.0\\
2013-10-25 &  300  & 56590.762 &  1.02 & 2.25  $\pm$ 0.22        &0.06     & 4.2\\
2013-10-26 & 300   & 56591.806 & 1.02  & 2.40 $\pm$ 0.24         &0.05     & 4.4\\
2013-10-27 & 600  & 56592.728 & 1.02  & 2.25 $\pm$ 0.18          & 0.10    & 4.2\\
2013-10-29 & 300  & 56594.803  &1.02  & 2.53 $\pm$ 0.41          &-0.12    & 4.5\\
2013-10-31 & 300  & 56596.705   &1.02  &2.67 $\pm$ 0.13          &-0.03    & 4.7\\
2013-11-17$^{b}$ & 900 & 56613.614  & 1.47 & 3.32 $\pm$ 0.13  &-0.06   & 5.4\\
2013-11-23$^{b}$ & 240 & 56619.633  & 1.47 & 3.02  $\pm$ 0.45  &-0.03  & 5.1\\
2014-09-17 & 1200 & 56917.793 &1.02  &5.64 $\pm$ 0.16             &0.05    &7.6\\
2014-09-29 & 600 & 56929.836  & 1.02 & 5.40 $\pm$ 0.06             &-0.03   & 7.4\\

\enddata
\label{table:log}
\tablenotetext{a:}{V/R=(I(V)-$I_c$)/(I(R)-$I_c$)}
\tablenotetext{b:}{The data are obtained with the Lijiang 2.4m telescope.}

\end{deluxetable}

\begin{deluxetable}{lcccc}
\tablecolumns{5} \tablecaption{Summary of the $BVRI$ photometric observations of LSV+44 17 between our 2007 and 2014 observations .}
\tablewidth{0pc}
\tablehead{\colhead{MJD}&\colhead{B (mag)} & \colhead{V (mag)} & \colhead{R (mag)} &\colhead{I (mag)}}  \startdata

54403.3002	&	-3.214$\pm$0.010	&	-3.511$\pm$0.010	&	--	&	--\\
54403.3895	&	-3.190$\pm$0.010	&	-3.493$\pm$0.010	&	--	&	--\\
54404.3290	&	-3.176$\pm$0.010	&	-3.476$\pm$0.010	&	--	&	--\\
54405.2250	&	-3.175$\pm$0.010	&	-3.484$\pm$0.010	&	--	&	--\\
54744.3119	&	-3.191$\pm$0.010	&	-3.482$\pm$0.010	&	--	&	--\\
54744.3177	&	-3.189$\pm$0.010	&	-3.485$\pm$0.010	&	--	&	--\\
54745.2810	&	-3.150$\pm$0.010	&	-3.440$\pm$0.010	&	--	&	--\\
54745.2863	&	-3.143$\pm$0.011	&	-3.440$\pm$0.010	&	--	&	--\\
54747.2616	&	-3.180$\pm$0.010	&	-3.462$\pm$0.011	&	--	&	--\\
54747.2666	&	-3.176$\pm$0.011	&	-3.454$\pm$0.012	&	--	&	--\\
54747.2703	&	-3.173$\pm$0.011	&	-3.459$\pm$0.012	&	--	&	--\\
54747.3697	&	-3.188$\pm$0.011	&	-3.462$\pm$0.012	&	--	&	--\\
54747.3746	&	-3.186$\pm$0.011	&	-3.471$\pm$0.012	&	--	&	--\\
54747.3794	&	-3.195$\pm$0.011	&	-3.469$\pm$0.013	&	--	&	--\\
54748.3333	&	-3.173$\pm$0.010	&	-3.459$\pm$0.010	&	--	&	--\\
54748.3394	&	-3.166$\pm$0.010	&	-3.457$\pm$0.010	&	--	&	--\\
55129.3571	&	-3.156$\pm$0.012	&	-3.442$\pm$0.012	&	--	&	--\\
55129.3580	&	-3.158$\pm$0.012	&	-3.450$\pm$0.012	&	--	&	--\\
55129.3589	&	-3.153$\pm$0.012	&	-3.443$\pm$0.012	&	--	&	--\\
55130.2832	&	-3.140$\pm$0.012	&	-3.430$\pm$0.012	&	--	&	--\\
55130.2841	&	-3.151$\pm$0.012	&	-3.430$\pm$0.012	&	--	&	--\\
55131.2841	&	-3.139$\pm$0.019	&	-3.455$\pm$0.018	&	--	&	--\\
55131.2850	&	-3.152$\pm$0.019	&	-3.433$\pm$0.018	&	--	&	--\\
55131.3613	&	-3.159$\pm$0.015	&	-3.453$\pm$0.015	&	--	&	--\\
55131.3622	&	-3.149$\pm$0.015	&	-3.454$\pm$0.014	&	--	&	--\\
55132.3163	&	-3.151$\pm$0.015	&	-3.440$\pm$0.014	&	--	&	--\\
55481.2514	&	-3.090$\pm$0.011	&	-3.349$\pm$0.011	&	-3.430$\pm$0.014	&	-2.076$\pm$0.012\\
55482.3408	&	-3.126$\pm$0.011	&	-3.364$\pm$0.011	&	-3.415$\pm$0.014	&	-2.074$\pm$0.011\\
55483.2688	&	-3.120$\pm$0.011	&	-3.362$\pm$0.011	&	-3.412$\pm$0.014	&	-2.085$\pm$0.011\\
55484.2269	&	-3.111$\pm$0.012	&	-3.358$\pm$0.012	&	-3.416$\pm$0.017	&	-2.085$\pm$0.011\\
55822.2986	&	-3.113$\pm$0.011	&	-3.345$\pm$0.010	&	-3.383$\pm$0.013	&	-2.017$\pm$0.011\\
55822.3428	&	-3.112$\pm$0.011	&	-3.346$\pm$0.010	&	-3.379$\pm$0.012	&	-2.008$\pm$0.011\\
55998.9640	&	--	&	-3.339$\pm$0.009	&	--	&	--\\
56216.2009	&	-3.081$\pm$0.012	&	-3.262$\pm$0.014	&	-3.271$\pm$0.015	&	-1.813$\pm$0.010\\
56216.2775	&	-3.096$\pm$0.012	&	-3.285$\pm$0.014	&	-3.275$\pm$0.014	&	-1.822$\pm$0.010\\
56216.3818	&	-3.079$\pm$0.014	&	-3.265$\pm$0.015	&	-3.219$\pm$0.014	&	-1.801$\pm$0.011\\
56217.2308	&	-3.083$\pm$0.011	&	-3.265$\pm$0.013	&	-3.241$\pm$0.013	&	-1.819$\pm$0.010\\
56217.3613	&	-3.114$\pm$0.011	&	-3.299$\pm$0.013	&	-3.262$\pm$0.013	&	-1.834$\pm$0.010\\
56217.3988	&	--	&	-3.290$\pm$0.015	&	-3.264$\pm$0.014	&	-1.827$\pm$0.010\\
56218.2771	&	-3.046$\pm$0.012	&	-3.261$\pm$0.013	&	-3.224$\pm$0.013	&	-1.796$\pm$0.010\\
56218.3386	&	-3.059$\pm$0.012	&	-3.264$\pm$0.013	&	-3.244$\pm$0.013	&	-1.817$\pm$0.010\\
56218.3571	&	-3.063$\pm$0.012	&	--	&	-3.247$\pm$0.013	&	-1.818$\pm$0.010\\
56219.2072	&	--	&	-3.258$\pm$0.017	&	-3.233$\pm$0.015	&	-1.801$\pm$0.011\\
56219.2783	&	-3.054$\pm$0.013	&	-3.258$\pm$0.015	&	-3.218$\pm$0.015	&	-1.801$\pm$0.011\\
56219.3524	&	-3.058$\pm$0.013	&	-3.270$\pm$0.015	&	-3.254$\pm$0.015	&	-1.815$\pm$0.011\\
56219.3939	&	--	&	-3.264$\pm$0.015	&	-3.244$\pm$0.015	&	-1.809$\pm$0.011\\
56221.2912	&	-3.073$\pm$0.012	&	-3.294$\pm$0.014	&	-3.268$\pm$0.014	&	-1.823$\pm$0.011\\
56221.3704	&	-3.038$\pm$0.012	&	-3.256$\pm$0.015	&	-3.226$\pm$0.018	&	-1.807$\pm$0.011\\
56222.2445	&	-3.046$\pm$0.012	&	-3.260$\pm$0.013	&	-3.224$\pm$0.013	&	-1.791$\pm$0.010\\
56222.3022	&	-3.069$\pm$0.012	&	-3.266$\pm$0.013	&	-3.253$\pm$0.013	&	-1.805$\pm$0.010\\
56222.3430	&	-3.062$\pm$0.012	&	-3.267$\pm$0.014	&	-3.231$\pm$0.013	&	-1.798$\pm$0.010\\
56222.3824	&	-3.045$\pm$0.012	&	-3.254$\pm$0.014	&	-3.228$\pm$0.013	&	-1.793$\pm$0.010\\
56558.3092	&	-3.061$\pm$0.013	&	-3.206$\pm$0.011	&	-3.184$\pm$0.011	&	-1.772$\pm$0.009\\
56558.3107	&	--	&	-3.247$\pm$0.010	&	-3.230$\pm$0.011	&	-1.806$\pm$0.009\\
56590.2584	&	-3.089$\pm$0.011	&	-3.256$\pm$0.013	&	-3.239$\pm$0.015	&	-1.862$\pm$0.012\\
56590.2057	&	-3.073$\pm$0.011	&	-3.278$\pm$0.013	&	-3.267$\pm$0.015	&	-1.855$\pm$0.012\\
56590.3155	&	-3.068$\pm$0.011	&	-3.275$\pm$0.013	&	-3.254$\pm$0.015	&	-1.862$\pm$0.012\\
56590.3705	&	-3.057$\pm$0.011	&	-3.269$\pm$0.013	&	-3.245$\pm$0.015	&	-1.844$\pm$0.012\\
56591.1350	&	-3.061$\pm$0.012	&	-3.271$\pm$0.014	&	-3.227$\pm$0.017	&	-1.852$\pm$0.013\\
56591.1910	&	-3.053$\pm$0.012	&	-3.242$\pm$0.015	&	-3.250$\pm$0.017	&	-1.840$\pm$0.013\\
56591.2440	&	-3.050$\pm$0.012	&	-3.253$\pm$0.015	&	-3.222$\pm$0.018	&	-1.834$\pm$0.013\\
56591.3051	&	-3.046$\pm$0.013	&	-3.250$\pm$0.015	&	-3.237$\pm$0.017	&	-1.841$\pm$0.013\\
56591.3678	&	-3.054$\pm$0.012	&	-3.266$\pm$0.015	&	-3.243$\pm$0.018	&	-1.838$\pm$0.013\\
56594.0885	&	-3.083$\pm$0.011	&	-3.281$\pm$0.011	&	-3.267$\pm$0.012	&	-1.859$\pm$0.010\\
56594.1659	&	-3.086$\pm$0.011	&	-3.277$\pm$0.010	&	-3.255$\pm$0.011	&	-1.854$\pm$0.009\\
56594.2230	&	-3.083$\pm$0.011	&	-3.261$\pm$0.010	&	-3.247$\pm$0.011	&	-1.847$\pm$0.009\\
56594.2844	&	-3.067$\pm$0.011	&	-3.253$\pm$0.010	&	-3.240$\pm$0.011	&	-1.837$\pm$0.009\\
56594.3180	&	-3.062$\pm$0.011	&	-3.258$\pm$0.010	&	-3.245$\pm$0.011	&	-1.833$\pm$0.009\\
56594.3427	&	-3.060$\pm$0.011	&	-3.255$\pm$0.010	&	-3.241$\pm$0.011	&	-1.833$\pm$0.009\\
56594.3776	&	-3.061$\pm$0.011	&	-3.254$\pm$0.010	&	-3.238$\pm$0.011	&	-1.832$\pm$0.009\\
56595.1477	&	-3.060$\pm$0.011	&	-3.255$\pm$0.011	&	-3.241$\pm$0.012	&	-1.835$\pm$0.010\\
56595.2087	&	-3.087$\pm$0.011	&	-3.270$\pm$0.011	&	-3.249$\pm$0.012	&	-1.855$\pm$0.009\\
56595.3489	&	-3.050$\pm$0.011	&	-3.248$\pm$0.011	&	-3.218$\pm$0.012	&	-1.817$\pm$0.010\\
56595.3850	&	-3.018$\pm$0.012	&	-3.222$\pm$0.011	&	-3.187$\pm$0.012	&	-1.778$\pm$0.010\\
56596.1719	&	-3.058$\pm$0.011	&	-3.249$\pm$0.010	&	-3.230$\pm$0.011	&	-1.837$\pm$0.009\\
56596.2621	&	-3.045$\pm$0.011	&	-3.245$\pm$0.011	&	-3.229$\pm$0.012	&	-1.828$\pm$0.010\\
56664.1353	&	-3.062$\pm$0.013	&	-3.259$\pm$0.012	&	-3.246$\pm$0.014	&	--\\
56685.0365	&	-3.035$\pm$0.015	&	-3.249$\pm$0.014	&	-3.238$\pm$0.016	&	-1.868$\pm$0.010\\
56697.0873	&	--	&	-3.323$\pm$0.017	&	-3.332$\pm$0.021	&	-1.901$\pm$0.012\\
56697.9922	&	--	&	-3.225$\pm$0.016	&	-3.256$\pm$0.020	&	-1.873$\pm$0.012\\
56719.9717	&	-3.037$\pm$0.013	&	-3.244$\pm$0.012	&	-3.261$\pm$0.014	&	-1.894$\pm$0.010\\
56724.9660	&	--	&	-3.275$\pm$0.015	&	-3.290$\pm$0.019	&	-1.914$\pm$0.010\\
56725.9744	&	--	&	-3.258$\pm$0.023	&	-3.257$\pm$0.026	&	-1.904$\pm$0.012\\
56926.3058	&	-3.102$\pm$0.012	&	-3.365$\pm$0.010	&	-3.403$\pm$0.012	&	-2.079$\pm$0.009\\
56928.3238	&	-3.007$\pm$0.017	&	--	&	-3.376$\pm$0.023	&	-2.050$\pm$0.012\\
56929.2070	&	-3.090$\pm$0.013	&	-3.379$\pm$0.010	&	-3.417$\pm$0.013	&	-2.093$\pm$0.009\\
56929.2953	&	-3.108$\pm$0.012	&	-3.375$\pm$0.010	&	-3.410$\pm$0.013	&	-2.087$\pm$0.009\\

\enddata
\label{table:logphot}

\end{deluxetable}

\end{document}